\documentclass[prb,preprint,superscriptaddress,nobibnotes]{revtex4}
\usepackage[dvips]{epsfig}
\usepackage{subfigure}
\usepackage{float}
\usepackage{amsmath}
\usepackage{amssymb}
\usepackage{amsfonts}
\usepackage{rotating}
\usepackage{euscript}
\usepackage{enumerate}
\usepackage{hhline}
\usepackage{threeparttable}
\usepackage{supertabular}
\usepackage{pslatex}
\usepackage{tabularx}
\usepackage{color}
\usepackage{hyperref}

\newcommand{\meff}{m_{\mathrm{eff}}}
\newcommand{\Vm}{V_\mathrm{m}}
\newcommand{\Vo}{V_\mathrm{o}}
\newcommand{\Leff}{L_{\mathrm{OM}}}

\newcommand{\num}{\nu_{\mathrm{m}}}
\newcommand{\nuo}{\nu_{\mathrm{o}}}
\newcommand{\Omegam}{\Omega_{\mathrm{m}}}
\newcommand{\omegao}{\omega_{\mathrm{o}}}

\newcommand{\modulus}[1]{\left|#1\right|}
\newcommand{\Imag}[1]{\mathrm{Im}\left\{#1\right\}}
\renewcommand{\Re}[1]{\mathrm{Re}\left\{#1\right\}}

\newcommand{\bv}[1]{\mathbf{#1}}
\newcommand{\maxvec}[1]{\textrm{max}\!\left( \left| \mathbf{#1} \right| \right)}

\begin{document}

\pagenumbering{arabic}

\title{Modeling Dispersive Coupling and Losses of Localized Optical and Mechanical Modes in Optomechanical Crystals}

\author{Matt Eichenfield}
\author{Jasper Chan}
\author{Amir H. Safavi-Naeini}
\author{Kerry J. Vahala}
\author{Oskar Painter}
\email{opainter@caltech.edu}
\homepage{http://copilot.caltech.edu}
\affiliation{Thomas J. Watson, Sr., Laboratory of Applied Physics, California Institute of Technology, Pasadena, CA 91125}

\begin{abstract}
Periodically structured materials can sustain both optical and mechanical excitations which are tailored by the geometry.  Here we analyze the properties of dispersively coupled planar photonic and phononic crystals: optomechanical crystals.  In particular, the properties of co-resonant optical and mechanical cavities in quasi-1D (patterned nanobeam) and quasi-2D (patterned membrane) geometries are studied.  It is shown that the mechanical $Q$ and optomechanical coupling in these structures can vary by many orders of magnitude with modest changes in geometry.  An intuitive picture is developed based upon a perturbation theory for shifting material boundaries that allows the optomechanical properties to be designed and optimized.  Several designs are presented with mechanical frequency $\sim 1$-$10$ GHz, optical $Q$-factor $Q_{o} > 10^7$, motional masses $\meff \approx$ 100 femtograms, optomechanical coupling length $\Leff < 5$ $\mu$m, and a radiation-limited mechanical $Q$-factor $Q_{m} > 10^7$.
\end{abstract}

\maketitle

\section{Introduction}\label{sec:Intro} It has previously been shown that ``defects'' in a planar periodic dielectric structure can simultaneously confine optical and mechanical resonances to sub-cubic-wavelength volumes\cite{ref:Maldovan_Simultaneous_Localization}.  As the co-localized resonances share the same lattice, and thus the same wavelength, the ratio of the optical to mechanical frequency of these modes is proportional to the ratio of their velocities.  More recently, it was demonstrated that such co-localized resonances in a Silicon structure can strongly couple, via motion-induced phase modulation of the internal optical field, resulting in sensitive optical read-out and actuation of mechanical motion at GHz frequencies\cite{ref:Eichenfield_OMC}.  In this paper we aim to further develop the theory and design of these coupled photonic and phononic systems, laying the groundwork for what we term ``optomechanical crystals''.  Here we choose a cavity-centric viewpoint of the interaction between photons and phonons, using the terminology and metrics from the field of cavity optomechanics~\cite{ref:TJK_KJV_review,ref:TJK_KJV_Science,ref:Favero09}.  An alternative viewpoint, more appropriate for guided-wave structures, may also be taken in which the interactions are described from a nonlinear optics (Raman-like scattering) perspective~\cite{ref:Trigo02,ref:Kang}.    

We focus on two cavity devices in particular, a quasi-one-dimensional (quasi-1D) patterned nanobeam and a quasi-two-dimensional (quasi-2D) patterned nanomembrane, both of which have been studied extensively in the past\cite{ref:Foresi,ref:Painter3} for their photonic properties.  The strength of the (linear) optomechanical coupling in such structures is found to be extremely large, approaching a limit corresponding to the transfer of photon momentum to the mechanical system every optical cycle\cite{ref:ChanJ1}.  Simultaneously, the effective motional mass\cite{ref:Pinard1} of the highly confined phonon modes is small, less than few hundred femtograms for a cavity system operating at a wavelength of $1.5$ $\mu$m and a mechanical frequency of $~2$ GHz.  This combination of parameters makes possible the optical transduction of high-frequency (multi-GHz) mechanical vibrations\cite{ref:Carmon3,ref:Kang,ref:Carmon4,ref:Trigo02} with near quantum-limited displacement sensitivity\cite{ref:Arcizet2,ref:Tittonen1}.  Additionally, dynamical back-action\cite{Braginsky92} between the photon and phonon fields can be used to dampen\cite{BraginskyTranquil,Gigan06,Arcizet06,Kleckner06,Schliesser06} and amplify\cite{Braginsky2001331,Kippenberg05,ref:Vahala_back_action_limit} mechanical motion, providing an optical source of coherent phonons\cite{ref:Wen,ref:lanzillotti-kimura} which can then be used within other phononic circuit elements\cite{ref:review_phononic_crystal_apps,ref:phononic_crystal_waveguide,ref:sound_attenuation_2d_array,ref:Robertson_acoustic_stop_bands,ref:psc_circuits_theory}.  Planar optomechanical crystals then, should enable a new generation of circuits where phonons and photons can be generated, routed, and made to interact, all on a common chip platform.

Unlike the simple motion of a mirror on a spring in more conventional cavity optomechanical systems\cite{ref:Dorsel,ref:Meystre}, the complex mechanics of optomechanical crystal structures makes it difficult to intuit the origin or strength of the optomechanical coupling.  Nonetheless, understanding the nature of the coupling is crucial to the engineering of optomechanical crystal devices as the degree of coupling between different optical and mechanical mode pairs can vary by orders of magnitude within the same structure, with even subtle changes in the geometry inducing large changes in the optomechanical coupling.  In the experimental demonstration of a nanobeam optomechanical crystal\cite{ref:Eichenfield_OMC}, it was shown that the perturbation theory of Maxwell's equations with shifting material boundaries~\cite{ref:Johnson_shifting_boundaries} provides an accurate method of estimating the optomechanical coupling of these complex motions.  Here we describe how this perturbation theory can be used to create an intuitive, graphical picture of the optomechanical coupling of simultaneously localized optical and mechanical modes in periodic systems.

The outline of the paper is as follows.  We first analyze the quasi-1D nanobeam optomechanical crystal system.  This nanobeam structure provides a simple example through which the salient features of optomechanical crystals can be understood.  The mechanical $Q$ of the structure is modeled using absorbing regions that provide a radiation condition for outgoing mechanical vibrations.  The various types of mechanical losses are analyzed, and methodologies for minimizing or avoiding these losses are discussed.  The dispersive coupling between the optical and mechanical modes is studied next.  We use the aforementioned perturbation theory to analyze the optomechanical coupling strength, and which we display as an optomechanical coupling density on the surface of the structure.  We use the density of optomechanical coupling picture to illustrate how the structure can be optimized to maximize the optomechanical coupling.  Finally, we analyze the optomechanical coupling of a quasi-2D membrane structure, the well-known double-heterostructure photonic crystal cavity~\cite{ref:Song}.  We show how the optical and mechanical modes and their coupling can be understood in terms of the quasi-one-dimensional nanobeam example.

\begin{figure*}[htb]
\begin{center}
\includegraphics[width=0.6\columnwidth]{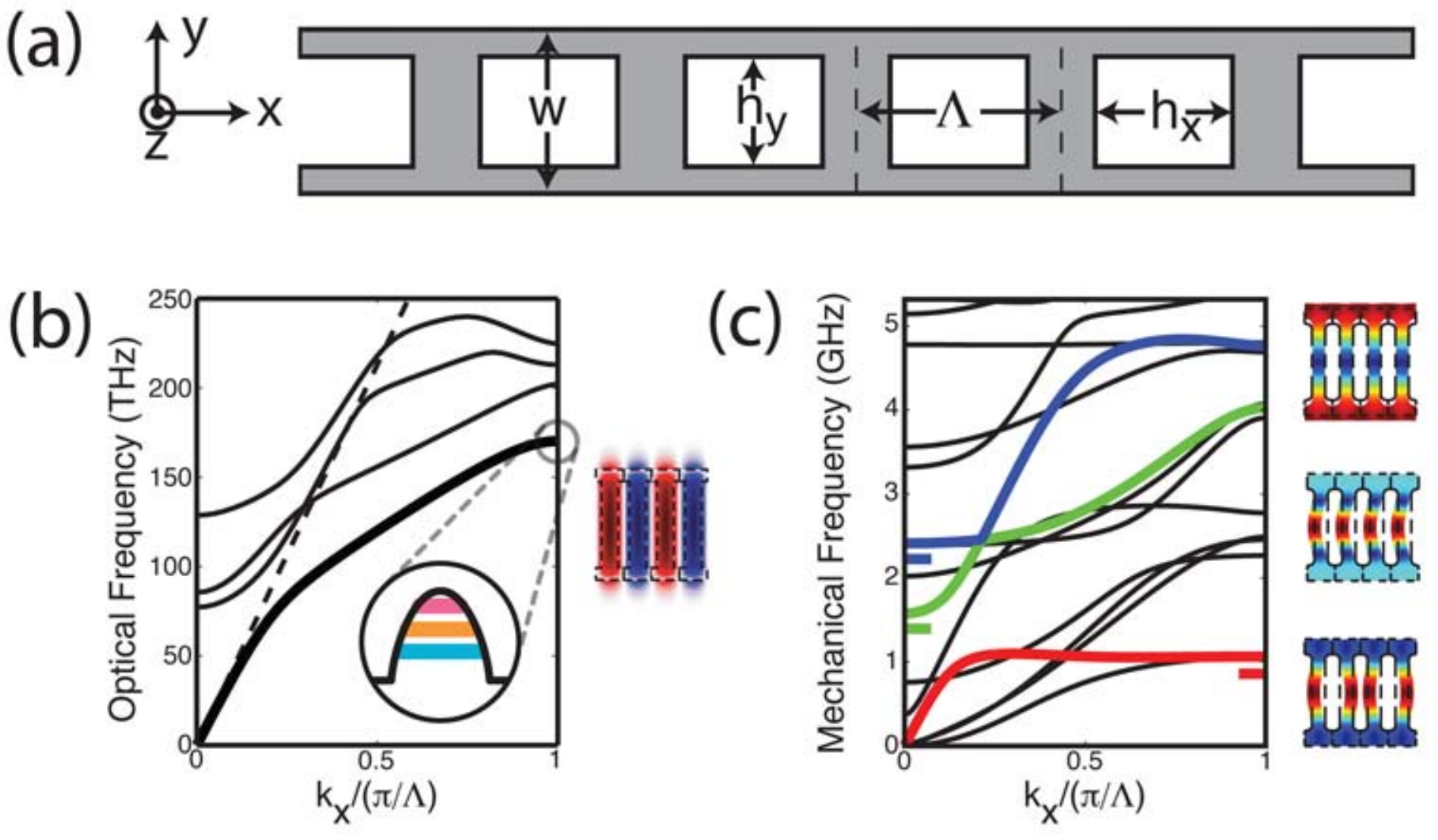}
\caption{\textbf{(a)}  General geometry of the periodic nanobeam structure's projection (infinite structure, no defect).  \textbf{(b)}  Optical band diagram of the nanobeam's projection.  The band from which all localized optical modes will be derived is shown in dark black, with $E_y$ of the optical mode at the $X$ point shown to the right of the diagram.  The harmonic spatial potential created by the defect, along with the first three optical modes are shown as emanating from the $X$-point band-edge.  \textbf{(c)}  Mechanical band diagram of the nanobeam's projection.  The three bands that form defect modes that will be discussed in this work are colored.  The bottom-most mode is from the $X$ point of the red band; the $\Gamma$ points of the green and blue bands correspond to the middle and top mechanical modes, respectively.  The frequencies of the defect modes that form from the band edges are shown as short, horizontal bars.} \label{fig:periodic}
\end{center}
\end{figure*}

% To generate the phonons, the optomechanical parametric instability\cite{Braginsky2001331,ref:Vahala_back_action_limit} can be used to create a self-sustaining mechanical oscillation with a continuous-wave laser input\cite{ref:Feng_oscillating_NEMS}, which provides a coherent source of phonons\cite{ref:Wen,ref:lanzillotti-kimura} over a large, technologically-relevant range of frequencies, with a low-noise, continuous-wave laser field acting as a pump for the output phonons.  These coherent sources of phonons, which can be routed to other on-chip phononic elements, could provide sufficiently large intensities to fully facilitate the study of nonlinear phononics, in much the same way that the laser paved the way for the study of nonlinear photonics\cite{ref:Franken_review}.  Because simple changes in the fabricated geometry can  vary the frequency of the mechanical mode from hundreds of megahertz to tens of gigahertz, optomechanical crystals provide a powerful, flexible platform for studying and controlling the interaction of light and mechanical vibrations, with the mechanical mode playing the role of a planar \emph{engineerable} Raman molecule\cite{ref:Kang,ref:Eichenfield_OMC,ref:Eichenfield_Zipper}. 

% , using optomechanical coupling to integrated planar lightwave circuits\cite{ref:Notomi2,ref:Song,ref:Painter3,ref:photonic_crystal_book} for all-optical detection, manipulation, and generation of the mechanical vibrations.  

\section{One-dimensional Optomechanical Crystal Systems:  An Example}\label{sec:quasi_1D} To illustrate the nature of the optomechanical coupling and losses in OMCs, we will use a quasi-1D nanobeam structure which has been demonstrated experimentally\cite{ref:Eichenfield_OMC}.  Figure~\ref{fig:periodic}(a) shows the general geometry of a periodic, quasi-one-dimensional OMC system made in a silicon beam of nanoscale cross-section.  The system consists of an infinitely periodic array of $h_x$ by $h_y$ rectangular holes with center-to-center spacing, $\Lambda$, in a beam of width $w$ and thickness, $t$ (not shown).  Although the actual structure will employ a defect to localize energy to a small portion of the beam, it is useful to consider the modes of this infinitely-periodic structure, since the structure has discrete translational invariance, allowing the optical and mechanical modes of the system to be classified according to their wavevector, $k_x$, and a band index.  We shall call the infinitely-periodic structure the \emph{projection} of the system.  The band picture provided by the projection allows a simple description of localized optical and mechanical modes as existing between two ``mirrors" in which propagating modes at the frequency of the defect have a small or vanishing density of states; the mirrors surround a perturbation region where propagation at the modal frequency is allowed, localizing the propagating mode between the mirrors.    The optical and mechanical bands of the OMC's projection are shown in Fig.~\ref{fig:periodic}(b) and \ref{fig:periodic}(c), respectively, for the structure $\Lambda = 360$ nm, $w = 1400$ nm, $h_y = 990$ nm, $h_x = 190$ nm, and $t = 220$ nm.  The material properties are parameterized by an isotropic Young's modulus, $E = 169$ GPa, and an index of refraction, $n = 3.49$.  The optical bands are computed with the MIT Photonic Bands package\cite{ref:MPB}, while the mechanical bands are computed with COMSOL Mutliphysics\cite{ref:COMSOL}, a finite element method (FEM) solver.  The structure does not possess a complete stop band for either the mechanics or the optics; nevertheless a defect in this structure can simultaneously produce highly-confined, low-loss optical and mechanical modes.

The primary optical mode of interest will be the first TE-like (dominantly polarized in the $y$-direction) "valence" band mode at the edge of the first Brillioun zone (the edge of the first Brillioun zone is called $X$ and the origin is called $\Gamma$).  The electric field profile, $E_y$, is shown next to the band diagram.  As described in detail in previous work on ``zipper'' optomechanical resonators\cite{ref:Eichenfield_Zipper,ref:ChanJ1} using general momentum-space design rules of photonic crystal cavities\cite{ref:Srinivasan1}, the localized modes that come from this band-edge mode are as far as possible from the light line while having a minimal amount momentum near $k_x = 0$ (and identically zero momentum at $k_x=0$) when used with a structure that is symmetric about a hole in the center.  This reduces the radiation loss out of the structure.  Because of the finite index contrast of the system, the optical $Q$ is limited by radiation from optical momentum components that are close to $k_x = 0$, since the system can only guide momentum components that are above the critical angle for total internal reflection.  This governs the design and choice of optical modes of the structure.  Because the optical band has negative curvature at the $X$ point, the frequency of the mode at the band edge must be increased to confine an optical mode coming from this band.  This can be accomplished by decreasing $\Lambda$ (making the holes closer together without changing the size of the hole).  As has been shown\cite{ref:Sauvan1,ref:Velha2,ref:Zain1,ref:McCutcheonM1,ref:Notomi7,ref:ChanJ1,ref:Eichenfield_Zipper,ref:Deotare09,ref:Loncar_high_Q_nanobeam} both theoretically and experimentally, these nanobeam systems are capable of achieving very high radiation-limited $Q$-factors.

Unlike light, mechanical energy cannot radiate into the vacuum.  This makes the design rules for creating low-loss mechanical defect modes qualitatively different than those discussed above for optical defect modes, as will be discussed in the next section.  Just as true \emph{photonic} band-gaps are not necessary to achieve high confinement and low optical losses in nanowire structures, true \emph{phononic} bandgaps are also unnecessary to achieve low mechanical losses.  A quasi-stop-band, where a defect mode of a particular polarization, frequency, and $k$-vector cannot couple a \emph{significant} amount of energy to the waveguide modes of the mirror portion will be enough to acheive mechanical energy localization.  Unlike in optics, all mechanical modes are ``guided" by the structure, regardless of their $k$-vector, which allows localized mechanical modes to be created from either of the high symmetry points, $\Gamma$ or $X$.  In fact, it will be shown that it is advantageous to draw mechanical modes from the $\Gamma$ point, as this generally produces larger optomechanical coupling than drawing from $X$.  Clearly the mechanical mode should be localized by the same defect as the optical mode; if this is not the case, then the target mechanical band-edge should be essentially unaffected by the defect that creates the optical mode, and a separate defect must be found that can localize the mechanics without significantly affecting the optical mode.  Finally, the localized defect mode that is formed from the band edge must be sufficiently optomechanically coupled to the localized optical mode(s) of interest.

\begin{figure*}[htb]
\begin{center}
\includegraphics[width=0.55\columnwidth]{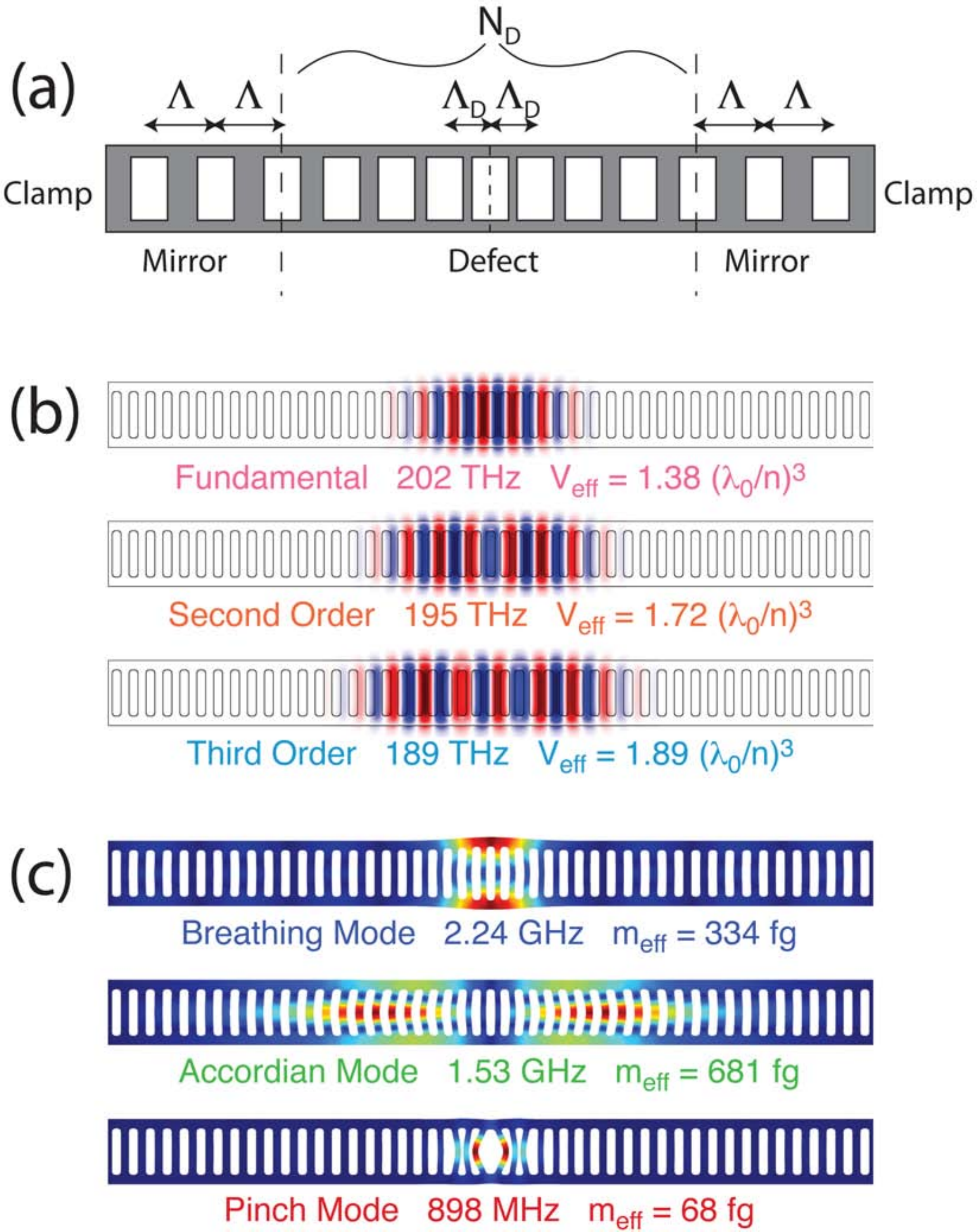}
\caption{\textbf{(a)} Schematic illustration of actual nanobeam optomechanical crystal with defect and clamps at substrate.  \textbf{(b)} Localized optical modes of the nanobeam OMC.  The colors of the names correspond to the illustration of the inverted potential in Fig.~\ref{fig:periodic}(b).  Localized, optomechanically-coupled mechanical modes of the nanobeam OMC.  The colors of the names correspond to the colored bands and horizontal bars showing the modal frequencies in Fig.~\ref{fig:periodic}(c).}  \label{fig:defect}
\end{center}
\end{figure*}

The defect that will be employed to localize optical and mechanical energy to the center of the structure consists of a decrease in the lattice constant for the otherwise periodic array of $N_{\mathrm{total}}$ holes in the beam, as illustrated in Fig.~\ref{fig:defect}(c).  For the modes that are localized by the defect, this effectively divides the structure into a "defect" portion where propagation is allowed, surrounded by ``mirrors", where the localized modes are evanescent, as discussed above.  The particular defect used here consists of some odd number of holes, $N_{\mathrm{D}}$, with the spacing between the holes varying quadratically from the background lattice constant, $\Lambda$, to some value $\Lambda_\mathrm{D}$, with the spacing varying symmetrically about the center hole (the hole dimensions are held fixed throughout the structure).    The complete geometry, which we will refer to as ``the nominal structure" is:  $N_{\mathrm{total}}= 75$, $\Lambda = 360$ nm, $w = 1400$ nm, $h_y = 990$ nm, $h_x = 190$ nm, $t = 220$ nm, $N_{\mathrm{D}}= 15$, and $\Lambda_\mathrm{D} = 0.85 \Lambda$.  In the nanobeam structure described here, this defect simultaneously localizes many mechanical and optical modes.

For localized modes, the quasi-harmonic spatial defect creates a quasi-harmonic potential for the optical and mechanical mode envelopes~\cite{ref:Painter14}.  This creates a ladder of states for each band edge with approximately Hermite-Gauss spatial dependencies along the cavity axis ($x$), in direct analogy to the harmonic potential of 1D quantum mechanics.  As discussed above, the localized optical modes of interest come entirely from a single band-edge (the darkened band in Fig.~\ref{fig:periodic}(b)); the first three cavity modes of the defect from that band are shown in Fig.~\ref{fig:defect}(b).  Many localized mechanical modes with linear optomechanical coupling exist in this system.  As examples, we will examine an exemplary optomechanically-coupled mechanical cavity mode from three different band-edges, even though each of these band edges produces a manifold of defect modes which may or may not have optomechanical coupling.  The three modes are shown in Fig.~\ref{fig:defect}(c), with the colors of the bands of Fig.~\ref{fig:periodic}(c) corresponding to the colors of the modes' label in the figure.

\begin{figure*}[htb]
\begin{center}
\includegraphics[width=0.625\columnwidth]{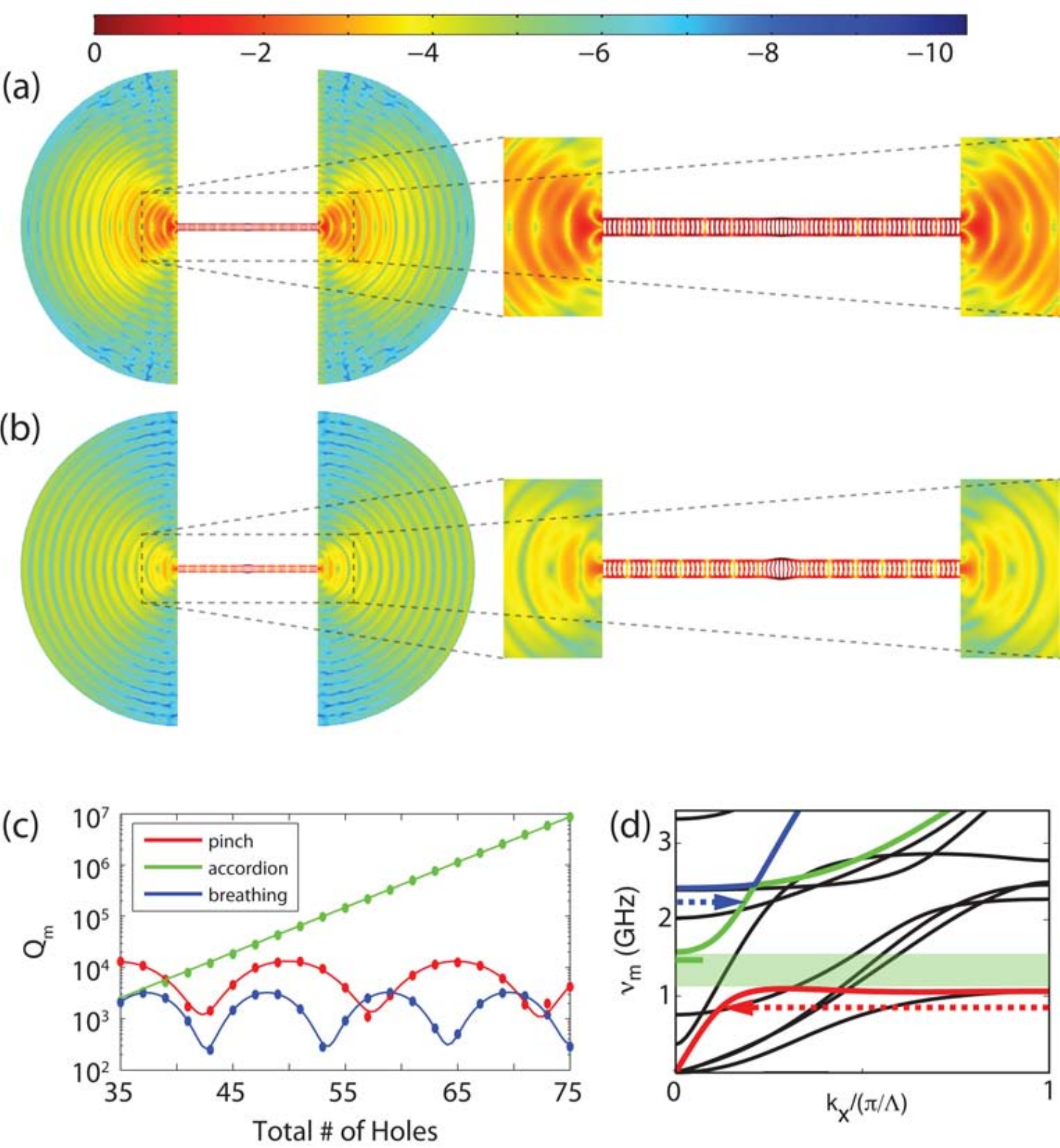}
\caption{\textbf{(a)} In-phase and \textbf{(b)} in-quadrature mechanical displacement field ($\log_{10}(| \bv{q}|^2/\textrm{max}(|\bv{q}|^2)$) of the fundamental breathing mode of nanobeam OMC structure with weakly absorbing ``pad", showing the propagating nature of the radiated mechanical waves in the pad region.  \textbf{(c)} Dependence of $Q_\textrm{m}$ on the total length of the structure; the number of mirror holes on each side is ($N_T-15)/2$.  This shows the oscillatory $Q_\textrm{m}$ of the pinch and breathing modes, which are coupled to waveguide modes, and the exponentially-increasing $Q_\textrm{m}$ of the accordion mode.  \textbf{(d)}  Mechanical band structure of the nanobeam OMC, with arrow tails indicating the frequency and high-symmetry point of the breathing (blue) and pinch (red) modes, and arrow heads indicating the equi-frequency waveguide mode that acts as the dominant source of parasitic coupling.  The effective bandgap of the accordion mode is shown in transparent green, with its frequency indicated as a horizontal green bar at the $\Gamma$ point.} \label{fig:Q}
\end{center}
\end{figure*}

\section{Modal Cross-Coupling and Mechanical Losses}\label{sec:mech_loss} Periodic structures can be fabricated to have phononic band gaps\cite{gorishnyy:121915,ref:Kushwaha_acoustic_bands,ref:Robertson_acoustic_stop_bands,ref:Espinosa_ultrasonic_band_gap}, where mechanical energy loss by linear elastic coupling to the environment can be made arbitrarily small.  Eventually, more fundamental losses\cite{ref:Cleland_book,ref:Kolsky_book} such as thermoelastic loss\cite{ref:Houston,ref:Yi,ref:Duwel,ref:Roukes_thermoelastic}, non-equilibrium energy redistribution\cite{ref:Landau_nonequilibrium_loss,ref:Kolsky_book}, phonon-phonon scattering\cite{ref:Fon_phonon-phonon_scattering}, and the movement of dislocations and impurities\cite{ref:Read_dislocation_friction,ref:Kolsky_book} should be accessible in these systems.  First, however, the linear interaction of the optomechanical crystal and its surrounding substrate, which acts as a bath, must be understood and minimized.  With this in mind, we use a finite element method model with weakly absorbing ``bath" regions to model the losses in the system due to coupling of the mechanical energy into modes that are not confined.  This method captures inter-modal coupling between the localized modes and all other mechanical modes of the system, some of which act as parasitic loss channels into the surrounding ``bath".

The lack of a mechanical bandgap means that the superposition of $k$-vectors necessary to create a localized mode in the defect coincide with $k$-vectors of equi-frequency propagating modes of the phononic crystal mirrors.  The localized modes and propagating modes of equal frequency will hybrizide and couple whenever the symmetries of the modes do not forbid it.  In addition to the propagating modes, there are ``body modes" that exist purely because of the boundary conditions (and thus not represented in the band structure), such as vibrations, density waves, and torsions of the finite, clamped structure.  If the simulated exterior boundary conditions allow energy in the propagating and body modes to be lost, the propagating and body modes that couple to the localized mode will act as parasitic loss channels for the localized mode.

In a fabricated structure, the cantilever is attached to a substrate at both ends, rather than a hard boundary at the end of the cantilever.  These more realistic boundaries must be included to model propagating and body mode losses.  The propagating modes travel down the nanobeam and partially reflect at the contacts due to an effective impedance mismatch caused by the geometric change between the nanobeam and the bulk.  The rest of the power radiates into the bulk, causing a loss of mechanical energy.   Thus the localized mode is coupled to a propagating mode with identical frequency that can radiate part or all of its energy into the surrounding ``bath".  This propagating mode also forms a coupled cavity resonance with the localized mode because of the reflections at the clamp points.  The body modes have a softer boundary condition than $\bv{q}=0$ at the boundaries, extending the body mode amplitude into the substrate.  The part of the body mode that extends into the substrate can excite radiative modes of the substrate.  The body mode then acts as a loss channel for any localized mode to which it is coupled.  The localized, propagating, and body modes form a set of coupled resonators.  Since the body and propagating modes are very sensitive to the total length of the structure, the self-consistent solution, which determines the loss of the localized mode, is very sensitive to the exact boundary conditions.  Thus, to accurately simulate the true spatial profiles and losses of localized mechanical modes, one needs a simulation that reflects the true boundary conditions.

To model the loss due to coupling to radiative modes of the substrate, we include a large, semi-circular ``pad" on each side of the nanobeam, with the same material constants as the nanobeam.  To make the pad act like a ``bath", we introduce a phenomenological imaginary part of the speed of sound in the pad region; i.e., $\textrm{v}_{\textrm{pad}} \rightarrow \textrm{v}_{\textrm{Silicon}}(1+i \eta)$, where $\textrm{v} = \sqrt{E/\rho}$.  This creates an imaginary part of the frequency, and the mechanical $Q$ can be found by the relation, $Q_\textrm{m} = \Re{\num}/(2 \Imag{\num})$.  By adding loss to the pad material, propagating modes will reflect part of their power at the contacts (the interface between the cantilever and the substrate) because of the change in the impedance from the absorption, not just the geometric change in impedance.  From this point of view, $\eta$ should be made as small as possible, since this contribution to the reflection coefficient is an artifact of the simulation and is not present in the real system.  However, $\eta$ must also be large enough that the self-consistent solution includes a propagating, radiated wave, which only happens if the wave is appreciably attenuated by the time it reflects from the edge of the simulation (where $\bv{q}=0$) and returns to the contact.  Thus, the pad is made as large as possible, given computational constraints, and the absorption is increased until $Q_\textrm{m}$ changes appreciably, which gives the threshold value for $\eta$ at which the reflectivity of the contacts has an appreciable contribution from the absorption.  The simulation is then performed with a value of $\eta$ that produces a propagating wave in the pad without causing an artificial reflectivity at the contact.  Propagation in the pad is easily verified if the position of the nodes/antinodes swap between the in-phase and in-quadrature parts of the mechanical cycle (the nodes/antinodes of a standing wave are stationary).  Figure \ref{fig:Q}(a) and \ref{fig:Q}(b) show the in-phase and in-quadrature (respectively) parts of the mechanical cycle of the breathing mode, clearly showing a propagating radiative mode in the weakly absorbing pad, with $\log_{10}(| \bv{q}|^2/\textrm{max}(|\bv{q}|^2)$ plotted to elucidate the attenuation of mechanical radiation in the pad. 

Limiting the artificial reflection at the interface of the non-absorbing and absorbing portions sets the maximum absorptivity, which in turn sets the minimum size of the pad (to guarantee that the radiation is completely attenuated before returning to the source).  This size/absorptivity trade-off can  be improved by making $\eta$ vary as a function of position in the pad, starting out at zero and increasing radially outward (quadratically, say).  This is analogous to a mechanical perfectly matched layer (PML)\cite{ref:Bindel_MPML}, which has the benefit of increasing the round-trip absorption while maintaining a minimum reflectance at the clamp \emph{due to} absorption.  This provides the same reflection-free absorption of mechanical radiation in a more compact simulation space, making better use of computational resources.

Changing the length of the structure changes the resonance condition for both the propagating and body modes.  This changes the amount of coupling to the localized mode in the self-consistent solution of the system.  Thus the parasitc losses into a waveguide mode should be periodic.  Figure \ref{fig:Q}(c) shows $Q_\textrm{m}$ for the pinch, accordion, and breathing mode as the total number of holes (i.e., the length of the optomechanical crystal) is varied.  For the breathing and pinch modes, the mechanical losses oscillate as a function of the total length of the nanobeam.  For this particular geometry, the losses are dominated by propagating modes, and the oscillation period of the $Q$ can be matched to a $k$-vector of a waveguide mode in the band structure in Fig. \ref{fig:Q}(d) (shown as a dotted line extending from the defect frequency).  Thus the length of the structure can be tuned to minimize mechanical losses in cases where a complete mechanical bandgap is not present.  Interestingly, the $Q$ of the accordion mode increases exponentially with the number of holes, indicating that the mode is evanescent in the mirror portions.  Examining the four bands that cross through at the frequency of the accordion mode, we find that all four bands have a mirror symmetry about either the $x-z$ or $x-y$ planes that forbids any hybridization or coupling between to the accordion mode.  This creates an effective bandgap (shown in translucent green).  Practically speaking, this is a much weaker stop-band than a true bandgap, because any defect in the structure that breaks the symmetry of the accordion mode about the mirror planes will cause a coupling to the waveguide modes in the gap.  However, it is exactly this kind of symmetry-dependent effective bandgap that is responsible for the high \emph{optical} $Q$\cite{ref:Eichenfield_Zipper,ref:Eichenfield_OMC, ref:Deotare09, ref:Loncar_high_Q_nanobeam} of the experimentally-fabricated structures.  This gives some confidence that it is possible to fabricate structures that are defect-free to the degree necessary to achieve high $Q_\textrm{m}$.

\section{Optomechanical Coupling:  definition and integral representation}\label{sec:OM_coupling_def} Cavity optomechanics involves the mutual coupling of two modes of a deformable structure:  one optical and one mechanical.  The optical mode is characterized by a resonant frequency $\omegao = 2 \pi \nuo$ and electric field $\bv{E}(\bv{r})$.  The mechanical mode is characterized by a resonant frequency $\Omegam = 2\pi \num$ and displacement field $\bv{Q}(\bv{r})$, where $\bv{Q}(\bv{r})$ is the vector displacement describing perpendicular displacements of the boundaries of volume elements.  The cavity optomechanical interactions of the distributed structure and its spatially-dependent vector fields, $\bv{E}(\bv{r})$ and $\bv{Q}(\bv{r})$, can be reduced to a description of two \emph{scalar} mode amplitudes and their associated mode volumes, with the coupling of the amplitudes parameterized by a single coupling coefficient, $g_{\mathrm{OM}}$. 

The mode amplitude, $c$, and complex vector field profile, $\bv{e}(\bv{r})$, are defined such that the complex electric field is $\bv{E}(\bv{r})=c\bv{e}(\bv{r})$ (the physical field is given by the real part of $\bv{E}(\bv{r})e^{i\omega t}$).  For pedagogical reasons, the amplitude $c$ is normalized such that the time averaged electromagnetic energy is equal to $\modulus{c}^2$; i.e. $U = \modulus{c}^2 = \frac{1}{2}\int \mathrm{d}V \epsilon \modulus{\bv{E}}^2$.  This forces $\bv{e}$ to be normalized such that $1= \frac{1}{2}\int \mathrm{d}V \epsilon \modulus{\bv{e}}^2$.  In cavity quantum electrodynamics, one typically defines an effective optical mode volume, $\Vo = \int \mathrm{d}V \left(\frac{ \sqrt{\epsilon}\modulus{\bv{E}}}{\maxvec{\sqrt{\epsilon} E}}\right)^2$, in order to gauge the strength of light-matter interactions.

The mechanical vibration's amplitude, $\alpha$, and mode profile (displacement), $\bv{q}(\bv{r})$, are defined such that $\bv{Q}(\bv{r})=\alpha\bv{q}(\bv{r})$.  Here, $\alpha$ is defined as the largest displacement that occurs anywhere for the mechanical field, $\bv{Q}(\bv{r})$, so that $\mathrm{max}(|\bv{Q}(\bv{r})|)=1$.  It is important to note that this particular choice of $\alpha$ determines the mechanical mode's effective volume and effective mass, $\Vm$ and $\meff \equiv \rho \Vm$, respectively.  In order to represent an energy and be consistent with the equipartition theorem, this choice of $\alpha$ requires the complimentary definition $\meff = \rho \int \mathrm{d}V \left(\frac{\modulus{\bv{Q}}}{\maxvec{Q}}\right)^2$.  To see this, note that the free evolution of the mechanical oscillator has, by definition, a time-independent total energy $E_{\textrm{mechanical}} = \frac{\meff}{2}(\Omega^2 \alpha^2 + \dot{\alpha}^2)$.  On the other hand, integrating the total energy of each volume element must \emph{also} give this same total energy.  If we pick the point in phase space at which all the mechanical energy is potential energy (i.e. the classical ``turn-around point"), we must have that $E_{\textrm{mechanical}} = \frac{1}{2}\Omega^2 \int \rho |\bv{Q}(\bv{r})|^2 \mathrm{d}V = \frac{1}{2}\meff \Omega^2 \alpha^2$, or, in other words, $\meff \alpha^2 = \int \rho |\bv{Q}(\bv{r})|^2 \mathrm{d}V$.  One can arbitrarily choose the definition of the amplitude or the mass, but choosing one determines the other.  Note that $\alpha$ is also the amplitude of zero-point motion of the canonical position operator in a quantized treatment. For a system like a localized mode of a phononic crystal defect cavity, where only a very small, localized portion of the total mass undergoes appreciable motion, the most sensible choice of the mass is the amplitude-squared weighted density integral, which, as stated above, is the choice of mass associated with $\alpha = \mathrm{max}(|\bv{Q}(\bv{r})|)$.

The optomechanical coupling affects the optical mode by tuning the optical resonant frequency as a function of displacement, $\omegao(\alpha)$; whereas the coupling affects the mechanical mode by applying a force, which is expressed as a gradient of the cavity energy, $\mathrm{d}\modulus{c}^2/\mathrm{d}\alpha$.   The optical resonant frequency is usually expanded in orders of the (small) displacement, $\alpha$ around some equilibrium displacement, $\alpha_0$.

\begin{equation}
\omegao(\alpha) = \omegao\Big|_{\alpha=\alpha_0} + (\alpha-\alpha_0) \frac{\mathrm{d}\omegao}{\mathrm{d}\alpha}\Big|_{\alpha=\alpha_0} + (\alpha-\alpha_0)^2 \frac{\mathrm{d^2}\omegao}{\mathrm{d}\alpha^2}\Big|_{\alpha=\alpha_0}+...
\end{equation}

In the case that the terms higher than first order can be neglected, this equation simplifies to

\begin{equation}
\omegao(\alpha) = \omegao\Big|_{\alpha=\alpha_0} + (\alpha-\alpha_0) \frac{\mathrm{d}\omegao}{\mathrm{d}\alpha}\Big|_{\alpha=\alpha_0} \equiv \omegao + (\alpha-\alpha_0)g_{\mathrm{OM}} \equiv \omegao + (\alpha-\alpha_0)\frac{\omegao}{\Leff} \;,
\end{equation}

\noindent where $\omegao \equiv \omegao\Big|_{\alpha=\alpha_0}$ is the equilibrium resonance frequency of the optical mode, $g_{\mathrm{OM}} \equiv \frac{\mathrm{d}\omegao}{\mathrm{d}\alpha}\Big|_{\alpha=\alpha_0}$ is the derivative of the resonance frequency of the optical mode evaluated at equilibrium, and $\Leff$ is the \emph{effective optomechanical length} of the system.  The effective length, $\Leff$, is a universal parameter that relates displacement to a change in optical frequency (i.e. $\alpha/\Leff = \delta \omegao/\omegao$).  From the definition, $\Leff^{-1} \equiv \frac{1}{\omegao}\frac{\mathrm{d}\omegao}{\mathrm{d}\alpha}\Big|_{\alpha=\alpha_0}=g_{\mathrm{OM}}/\omega_0$, one can see that reducing $\Leff$ maximizes the optomechanical coupling.  It is simple to show that $\Leff$ is equal to the spacing between the mirrors of a Fabry-Perot cavity when one mirror is allowed to move along the cavity axis or the radius of a microtoroid/microdisk for a radial breathing motion.  For a ``Zipper" cavity or double-microdisk, $\Leff$ is an exponentially decreasing function of the spacing between the coupled elements, with $\Leff$ approaching half a wavelength of light as the spacing approaches zero.

The perturbation theory of Maxwell's equations with shifting material boundaries\cite{ref:Johnson_shifting_boundaries} allows one to calculate the derivative of the resonant frequency of a structure's optical modes, with respect to some parameterization of a surface deformation \emph{perpendicular} to the surface of the structure.  If the result of a mechanical simulation is the displacement field, $\bv{Q}(\bv{r}) = \alpha \bv{q}(\bv{r}) \equiv \alpha \bv{Q}(\bv{r})/\maxvec{Q}$, then

\begin{equation}\label{eq:Leff}
\frac{1}{\Leff} = \frac{1}{4}\displaystyle \int \mathrm{d}A \left(\bv{q} \cdot \bv{\hat{n}} \right) \left[\Delta \epsilon \modulus{\bv{e}_\parallel}^2 - \Delta (\epsilon^{-1}) \modulus{\bv{d}_\perp}^2 \right]
\end{equation}

\noindent where $\bv{d} = \epsilon \bv{e}$, $\bv{\hat{n}}$ is the unit normal vector on the surface of the unperturbed cavity, $\Delta \epsilon = \epsilon_1 - \epsilon_2$, $\Delta (\epsilon^{-1}) = \epsilon_1^{-1} - \epsilon_2^{-1}$, $\epsilon_1$ is the dielectric constant of the structure, and $\epsilon_2$ is the dielectric constant of the surrounding medium.

To calculate $\Leff$ by deforming the structure, one must simulate the fields with a deformation amplitude, $\alpha$, that is large enough to be detectable numerically but small enough that higher order dispersion does not affect the frequency shift.  To verify that higher order dispersion is not included, one must simulate the optical fields for a range of displacement amplitudes and extract the linear dispersion.  Because perturbation theory can calculate the linear term exactly from a single calculation using the \emph{undeformed} structure, this method has clear advantages over numerical methods using finite deformations.

\begin{figure*}[htb]
\begin{center}
\includegraphics[width=0.55\columnwidth]{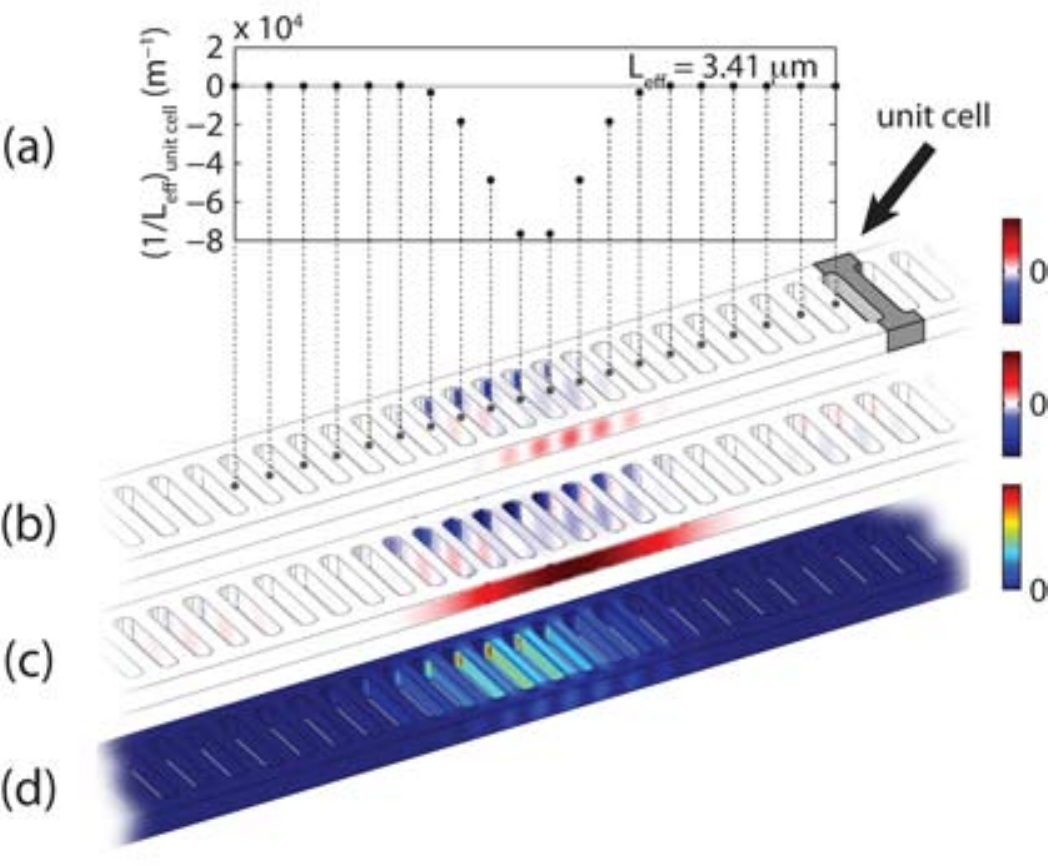}
\caption{For the fundamental breathing mode and the fundamental optical mode in the nominal structure, \textbf{(a)} FEM simulation of individual unit cell contributions to the total optomechanical coupling (each point computed by integrating $\zeta_{\mathrm{OM}}$ (Equation~\ref{eq:Leff}) over the respective unit cell), \textbf{(b)} surface plot of the optomechanical coupling density, $\zeta_{\mathrm{OM}}$.  \textbf{(c)} surface plot of the normal displacement profile, $\Theta_\textrm{m}$ (Equation~\ref{eq:dfunc}), \textbf{(d)} surface plot of the electromagnetic energy functional, $\Theta_\textrm{o}$ (Equation~\ref{eq:efunc}). In  (d), there is significant optomechanical coupling density in the corner of the holes, where the crossbar meets the rail. Without the fillets, the field amplitude is concentrated in the corner and difficult to see.  For this reason, the corners have been filleted to allow the optomechanical coupling density in the corners to be visualized.  The fillets do not significantly affecting the optomechanical coupling (confirmed by simulation). } \label{fig:visbreathe}
\end{center}
\end{figure*}

\section{Optomechanical coupling:  visual representation and optimization}\label{sec:OM_coupling_opt} In addition to being computationally simpler than deformation methods, the perturbative method of calculating the optomechanical coupling allows one to represent the optomechanical coupling as a density on the surface, with different parts of the structure contributing different amounts of optomechanical coupling.  This yields much more information than just the value of $\Leff$, itself. The optomechanical coupling density is given by

\begin{equation}\label{eq:Leffintegrand}
\zeta_{\mathrm{OM}}(\bv{r}) \equiv \frac{1}{4}{ \displaystyle \left(\bv{q} \cdot \bv{\hat{n}} \right) \left[\Delta \epsilon \modulus{\bv{e}_\parallel}^2 - \Delta (\epsilon^{-1}) \modulus{\bv{d}_\perp}^2 \right]}\;.
\end{equation}

The optomechanical coupling density can further be broken down into a mechanical part (the normal displacement profile)

\begin{equation}\label{eq:dfunc}
\Theta_\textrm{m}(\bv{r}) \equiv \bv{q} \cdot \bv{\hat{n}}
\end{equation}

\noindent and an optical part (the electromagnetic energy functional)

\begin{equation}\label{eq:efunc}
\Theta_\textrm{o}(\bv{r}) \equiv \Delta \epsilon \modulus{\bv{e}_\parallel}^2 - \Delta (\epsilon^{-1}) \modulus{\bv{d}_\perp}^2 \;,
\end{equation}

\noindent which can be separately visualized on the surface.  This provides a quantitative method of assessing the separate optical and mechanical contributions and allows an intuitive approach to individually engineering the optical and mechanical properties of the structure to enhance the optomechanical coupling of specific modes.

Figure~\ref{fig:visbreathe}(a) shows the contribution to the optomechanical coupling, $\Leff^{-1}$, of the breathing mode and fundamental optical mode from each ``unit cell" of the structure.  Summing the contributions from each unit cell yields  $\Leff^{-1}$.  Figs.~\ref{fig:visbreathe}(b)-(d) show $\zeta_{\mathrm{OM}}$, $\Theta_\textrm{m}$, and $\Theta_\textrm{o}$, plotted on the surface of the nanobeam OMC for the fundamental breathing mode and the fundamental optical mode.  In Fig.~\ref{fig:visbreathe}(b), it can be seen that there are two dominant and opposite contributions to the optomechanical coupling: one from the outside face of the rails and one from the inside face of the rails (in the corners of holes).  Minimizing the cancellation between these two contributions is critical to achieving a small $\Leff$ for the breathing mode (i.e. strong optomechanical coupling).  The geometry of the ``nominal" structure optimizes the coupling between the fundamental optical mode and the breathing mode, as shown below.

\begin{figure*}[htb]
\begin{center}
\includegraphics[width=0.55\columnwidth]{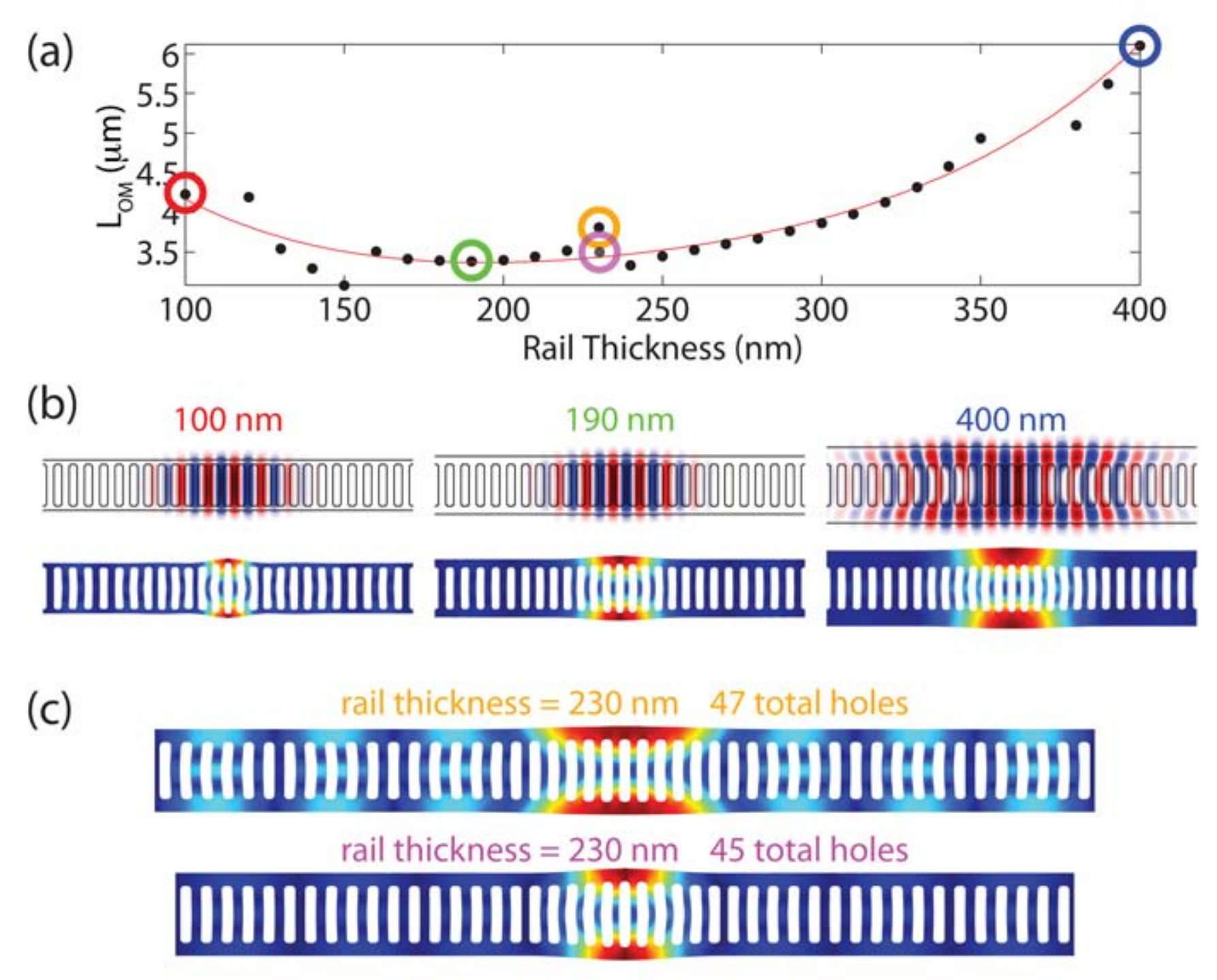}
\caption{For the fundamental breathing mode and the fundamental optical mode, \textbf{(a)} the dependence of the optomechanical coupling on the rail thickness (with oscillations in the data arising from accidental degeneracies with the cantilever modes), \textbf{(b)} the optical and mechanical mode profiles for rail thicknesses of 100~nm, 190~nm and 400~nm circled in red, green and blue respectively in (a), \textbf{(c)} comparison of the mechanical mode profiles when coupled (orange) and not coupled (purple) to cantilever modes, with the corresponding effect on  in $\Leff$ highlighted in (a). } \label{fig:rail_var}
\end{center}
\end{figure*}

Since the breathing mode is drawn from a band edge at the $\Gamma$ point, adjacent unit cells are mechanically in-phase with each other and add constructively to the optomechanical coupling.  This is in contrast to defect modes drawn from band edges at the $X$ point, such as the pinch mode, where adjacent unit cells are mechanically out-of-phase, resulting in neighboring unit-cell contributions that tend to cancel.  This cancellation reduces the optomechanical coupling unless it is specifically mitigated with extremely tight modal envelopes (see description of pinch mode optomechanical coupling below).

The degree to which the different faces of the rails cancel each other's contribution to $\Leff^{-1}$ is set by the attenuation of the optical field between the two edges, as the mechanical displacement of the two rails is fairly uniform.  Thus, one would expect that varying the rail thickness, which changes the relative amplitude of the optical field on the two rail faces, would have a significant impact on the coupling.  Figure~\ref{fig:rail_var}(a) shows $\Leff$ as a function of rail thickness, with $\Leff$ of the nominal structure (190~nm rail thickness) circled in green.  For rail thicknesses smaller than 190~nm (such as the 100~nm rail width, circled in red in Fig.~\ref{fig:rail_var}(a) and shown in Fig.~\ref{fig:rail_var}(b)), the amplitude of the optical field on the inside and outside edge of the field is becoming more and more similar.  This results in a larger cancellation between the contributions to $\Leff^{-1}$ on the inside and outside of the rails, decreasing the optomechanical coupling.  This reasoning might lead one to believe that increasing the rail thickness should monotonically decrease $\Leff$ (increase optomechanical coupling).  However, for rail thicknesses larger than 190~nm (such as the 400~nm rail width, circled in blue in Fig.~\ref{fig:rail_var}(a) and shown in Fig.~\ref{fig:rail_var}(b)), there is significant decrease in confinement of the optical mode because the light can partially ``spill around" the holes through the wide rails.  The mechanical mode, in contrast, stays relatively confined.  The net effect is that the optical energy is ``wasted" on parts of the structure that do not have significant motion, and the optomechanical coupling is again decreased. 

\begin{figure*}[htb]
\begin{center}
\includegraphics[width=0.55\columnwidth]{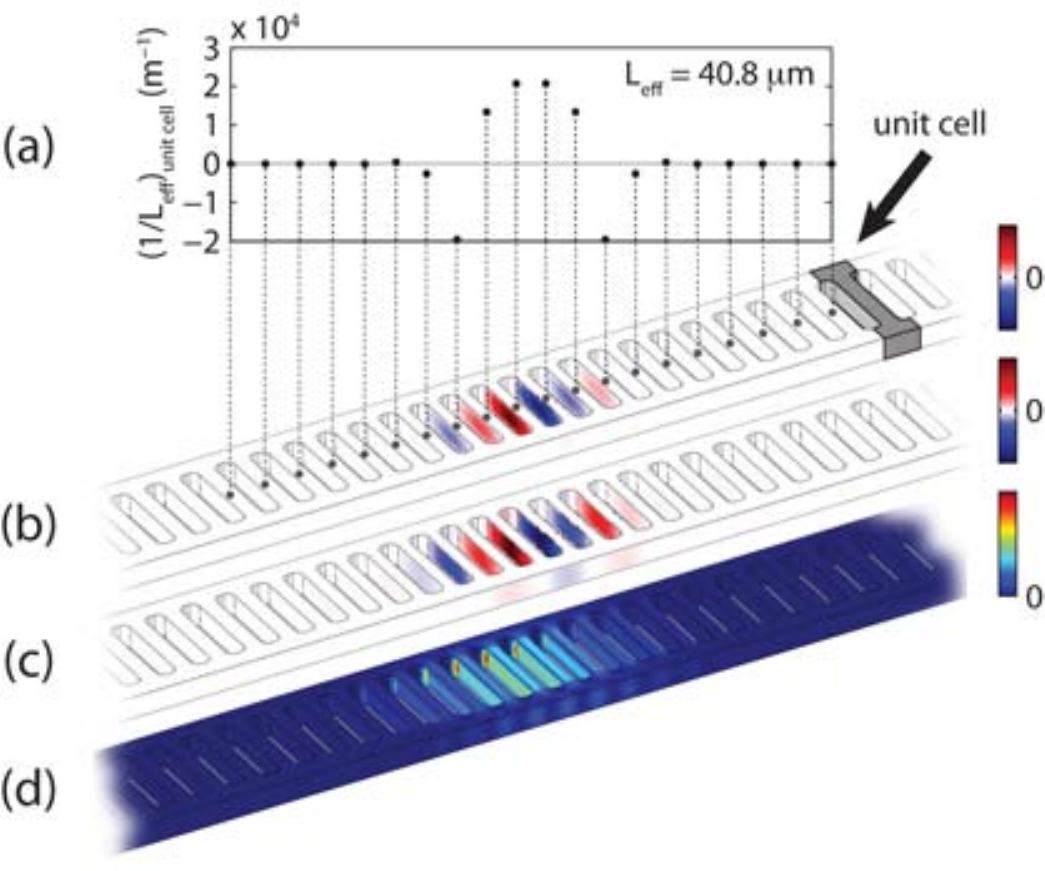}
\caption{For the fundamental pinch mode and the fundamental optical mode in the nominal structure, \textbf{(a)} FEM simulation of individual unit cell contributions to the total optomechanical coupling, \textbf{(b)} surface plot of the optomechanical coupling density, \textbf{(c)} surface plot of the normal displacement profile (Equation~\ref{eq:dfunc}), \textbf{(d)} surface plot of the electromagnetic energy functional (Equation~\ref{eq:efunc}).} \label{fig:vispinch}
\end{center}
\end{figure*}

Just as $Q_\textrm{m}$ is affected by hybridization of the breathing mode with propagating and body modes, $\Leff$ is affected by hybridization as $\bv{q}(\bv{r})$ is modified by the coupling to waveuide or body modes.  This is responsible for the oscillations in $\Leff$ seen in Fig.~\ref{fig:rail_var}(a).  The impact of coupling to the nanobeam body modes can be clearly seen in Fig.~\ref{fig:rail_var}(c), where the breathing mode in a structure with a rail thickness of 230~nm has been plotted for two different beam lengths (number of total holes). For 47 total holes (circled in orange in Fig.~\ref{fig:rail_var}a)), the breathing mode shape is altered significantly by the hybridization, causing the $\Leff$ to deviate from the trend indicated by the red line in Fig.~\ref{fig:rail_var}a). Shortening the structure by 2 holes (one on each side) decreases the coupling of the breathing mode to the propagating mode, returning $\Leff$ to the trend line.

\begin{figure*}[htb]
\begin{center}
\includegraphics[width=0.55\columnwidth]{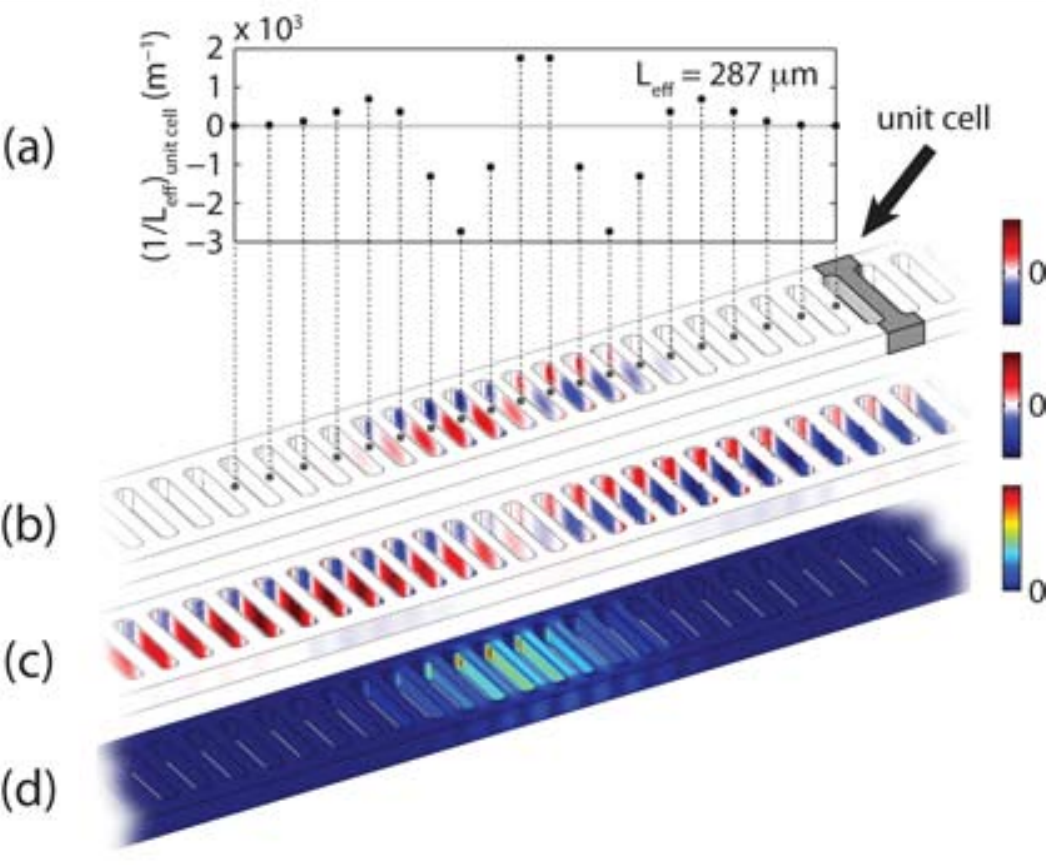}
\caption{For the accordion mode and the fundamental optical mode in the nominal structure, \textbf{(a)} FEM simulation of individual unit cell contributions to the total optomechanical coupling, \textbf{(b)} surface plot of the optomechanical coupling density, \textbf{(c)} surface plot of the normal displacement profile (Equation~\ref{eq:dfunc}), \textbf{(d)} surface plot of the electromagnetic energy functional (Equation~\ref{eq:efunc}).} \label{fig:visaccordion}
\end{center}
\end{figure*}

\begin{figure*}[htb]
\begin{center}
\includegraphics[width=0.55\columnwidth]{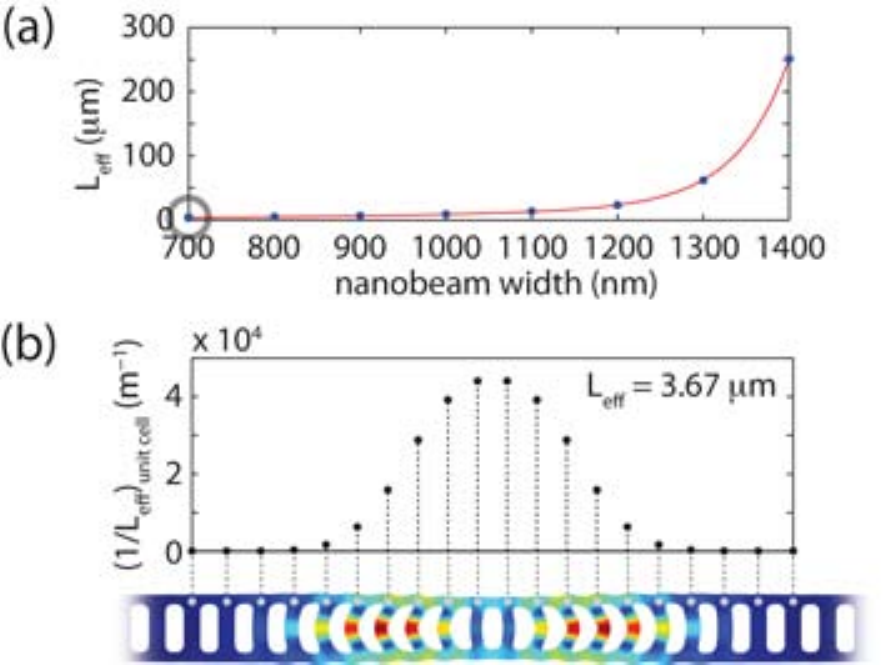}
\caption{For the accordion mode with the fundamental optical mode, \textbf{(a)}, the effective length as a function of total beam width, \textbf{(b)}, individual unit cell contributions to the total optomechanical coupling for a structure with a beam width of 700~nm (circled in (a)), mode frequency of 3.97~GHz and effective motional mass of 334~fg, with accompanying mechanical mode plot. The narrower mechanical mode (represented here by the deformation of the structure with color indicating relative strain) envelope results in drastically different optomechanical coupling contributions compared to Fig.~\ref{fig:visaccordion}.} \label{fig:visaccordion2}
\end{center}
\end{figure*}

The pinch mode is a localized, in-plane differential acoustic vibration.   Each neighboring crossbar vibrates 180 degrees out of phase with its nearest neighbors, since the pinch mode is drawn from a band edge at the $X$ point.  So although the optomechanical coupling contribution from each half of the structure (with respect to the $y$-$z$ plane) is equal, such that the two halves add constructively to $\Leff^{-1}$, on either side of the $y$-$z$ plane, contributions to $\Leff^{-1}$ from neighboring crossbars tend to cancel.  This puts a premium on mechanical localization, as a more localized pinch mode has a larger difference (and thus a reduced cancellation) between neighboring crossbars.  Although the envelope of the pinch mode's displacement profile is gaussian, each crossbar is very rigid, so the displacement of the compression and tension faces of each beam is essentially identical (but opposite).  The gaussian envelope only serves to change the relative vibration amplitudes of neighboring crossbars.  The optomechanical coupling contribution from each beam would be approximately \emph{zero} if it weren't for the rapid variation of the optical mode's envelope, and the contribution of each crossbar to $\Leff^{-1}$ depends primarily on the difference in the optical energy density across the width of the beam.  This would then lead one to believe that tighter localization, both optically and mechanically, would produce better optomechanical coupling for this structure.  Indeed, although the $\Leff$ of the pinch mode in the structure shown is quite modest ($\approx 41$ $\mu$m), $\Leff$ can be reduced to less than 3 $\mu$m by more tightly confining the optical and mechanical modes by reducing the number of holes involved in the defect region.  There is, however, a loss of optical $Q$ associated with the increased confinement due to the larger optical momentum components associated with tighter spatial localization.  However, the structure as shown has a radiation-limited optical $Q$ greater than 10 million; so it can be quite reasonable to trade optical $Q$ for higher optomechanical coupling.

The last type of mechanical mode to be considered is the accordion mode (Fig.~\ref{fig:visaccordion}). The relatively poor $\Leff$ for the accordion mode in the nominal structure is partly due to the fact that the rails recoil against the motion of the cross bar, producing opposing optomechanical contributions within each unit cell. In addition, the coupling of the broad first order Hermite-Gauss envelope of the mechanical mode with the narrower optical mode induces cancellations in the optomechanical coupling contributions at the inflection points of the mechanical mode envelope.   

As discussed above, the accordion mode has a large effective mechanical bandgap.  The dramatically increased $Q_\textrm{m}$ that results makes it worthwhile to investigate whether the structure can be modified to produce smaller $\Leff$.  By reducing the width of the nanobeam, it can be seen from Fig.~\ref{fig:visaccordion2}(a) that the coupling is dramatically improved by almost two orders of magnitude when the width of the structure is reduced.  As shown in Fig.~\ref{fig:visaccordion2}(b), for a beam width of 700~nm, the contributions within each unit cell no longer cancel, due to the comparatively narrower mechanical mode envelope, and the structure yields $\Leff = 3.67$ $\mu$m.  In addition, simulations of the $Q_\textrm{m}$ show that the effective bandgap for this narrower structure is approximately 2 GHz, yielding an extremely large $Q_\textrm{m}$ for a given number of holes in the mirror section ($Q_\textrm{m} \approx 10^8$ for 35 total holes).  It should be noted that the frequency of the accordion mode of the narrower structure is approximately 4 GHz, up from 1.5 GHz in the wider structure.

\begin{figure*}[ht]
\begin{center}
\includegraphics[width=0.525\columnwidth]{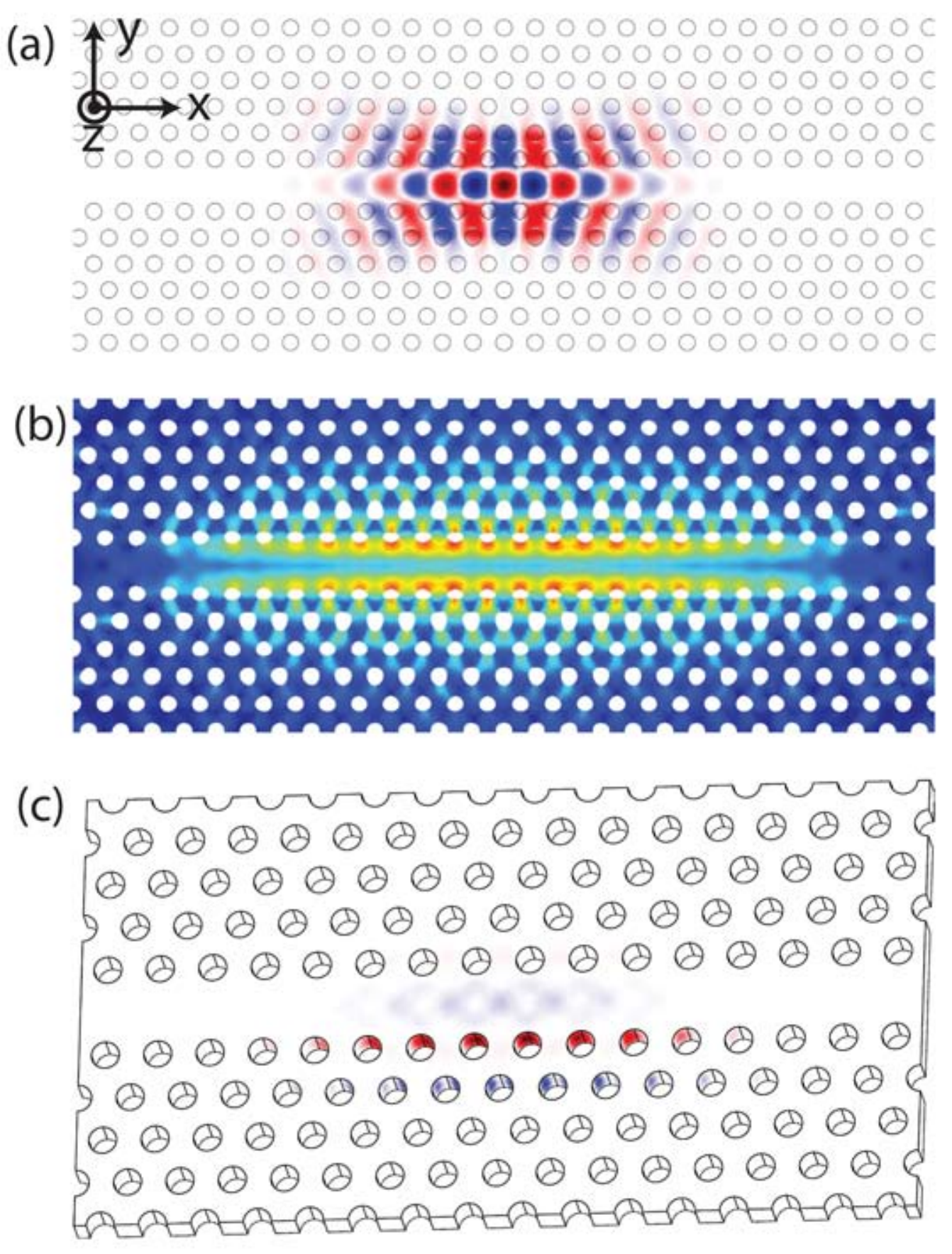}
\caption{\textbf{(a)} Fundamental optical mode of the double-heterostructrure OMC (geometry identical to that described in Ref.~\cite{ref:Song}), with $\lambda_0 \approx 1.5$ $\mu$m, $Q_{\mathrm{rad}}\approx 2.7\times 10^7$, and $V_{\mathrm{eff}}=1.2$ $(\lambda_0/n)^3$.  \textbf{(b)} Breathing mechanical mode of the double-heterostructure OMC, with $\nu_\textrm{m} = 9.3$ GHz, and $\meff = 322$ femtograms.  \textbf{(c)}  Optomechanical coupling integrand plotted on the double-heterostructure OMC system's surface; the structure has an $\Leff = 1.75$ $\mu$m for the optical-mechanical mode-pair from \ref{fig:hetero}(a) and \ref{fig:hetero}(b). } \label{fig:hetero}
\end{center}
\end{figure*}

\section{Optomechanical coupling in two-dimensional optomechanical crystals}\label{sec:quasi_2D_OMC} As a final example of how these methods can be used to understand the optomechanical coupling in periodic structures with complex mechanical and electric field profiles, we model a double heterostructure hexagonal photonic crystal slab resonator.  This is a well-known optical system, which has been found to have radiation-limited quality factors in excess of twenty million, with experimental demonstrations exceeding quality factors of two million\cite{ref:Song}.  The system consists of a hexagonal lattice of air holes in a silicon slab, with a single row of holes removed to create a waveguide mode within the optical stop band (the defect pulls the waveguide mode from the conduction band); in addition, the spacing in the direction of the waveguide is abruptly decreased twice to provide longitudinal confinement.  This structure is essentially equivalent to the nanobeam structure, with the optical and mechanical modes guided by Bragg reflection in the lateral direction, as opposed to total internal reflection and hard boundaries in the nanobeam.  With this in mind, we expect very similar optical and mechanical modes wherever lateral propagation out of the waveguide is prohibited by Bragg reflection. 

The fundamental optical cavity mode of the structure (the geometry is identical to that described in Ref.~\cite{ref:Song}) has been reproduced by FEM simulation and shown in Fig.~\ref{fig:hetero}(a).  The structure also exhibits a lateral mechanical breathing mode at 9.3 GHz with a motional mass of 300 femtograms, modulating the width of the waveguide in a way that is analogous to the mechanical breathing mode of the nanobeam.  The breathing mode displacement profile is shown in Fig.~\ref{fig:hetero}(b). 

Figure~\ref{fig:hetero}(c) shows the integrand of the optomechanical coupling integrand (Eq.~\ref{eq:Leff}) between the the optical mode and the mechanical breathing mode plotted on the surface of the structure.  The structure is shown slightly tilted to allow the insides of the holes to be seen, which give the dominant contributions to the optomechanical coupling.  It is interesting to note that the coupling comes almost entirely from the movement of a small part of the interior of the holes (i.e., the region of the inner sidewall of the hole, closest to the center defect region); this can be seen by comparing the top half of the structure to the bottom half (since the integrand is symmetric about the $x$-$z$ plane).  Since each row of holes provides an opposite contribution to its neighbors, it is necessary to have a rapidly decaying optical envelope to achieve small $\Leff$, which is the case for the optical mode shown here.  There is also a very small, opposing contribution from the center waveguide due to buckling/extrusion of the structure as the width is modulated.  Just as in the case of the nanobeam, this optical-mechanical mode-pair has a very strong dispersive coupling, and evaluating the integral yields an effective length of only 1.75 $\mu$m.

Optically, the structure has a complete photonic bandgap for in-plane propagation, but, with a hole size to lattice constant ratio of $r/\Lambda = 0.26$, there is no corresponding in-plane mechanical bandgap.  This makes the structure susceptible to mechanical loss mechanisms similar to those of the nanobeam.  However, the two-dimensional hexagonal lattice, as well as other two dimensional Bravais lattices, can have simultaneous optical and mechanical bandgaps\cite{ref:Thomas_simultaneous_gaps,ref:Mohammadi}, allowing the possibility of highly localized, low-loss optical-mechanical mode-pairs with very small effective lengths and motional masses. 

\section*{Acknowledgements}\label{sec:Ack} This work was funded through the NSF under EMT grant no. 0622246, and MRSEC grant no. DMR-0520565, and CIAN grant no. EEC-0812072 through University of Arizona.

%\bibliography{../MEbib_20090729_OJP}

\end{document}